\documentclass[10pt,conference]{IEEEtran}
\usepackage{epsfig,epsf}
\usepackage{color}
\usepackage{bbold}
\usepackage{multirow}
\usepackage{latexsym}
\usepackage{amssymb}
\usepackage{amsmath}
\usepackage{amsfonts}
\usepackage[sort]{cite}
\usepackage{mathrsfs}
\usepackage{scalefnt}
\usepackage{extarrows}
\usepackage{setspace}
\usepackage{stfloats}
\usepackage{url}
\usepackage{hyphenat}
\usepackage{graphicx}
\usepackage[table]{xcolor}
\usepackage{bbm}
\usepackage{dsfont}
\usepackage{bm}
\usepackage{array}
\usepackage{wrapfig}
\usepackage[]{algpseudocode}
\usepackage{algorithm}

\usepackage{mathtools}
\allowdisplaybreaks[1]
\usepackage{epstopdf}
\newcommand\numberthis{\addtocounter{equation}{1}\tag{\theequation}}
\pdfpageattr {/Group << /S /Transparency /I true /CS /DeviceRGB>>}

\makeatletter
\newsavebox\myboxA
\newsavebox\myboxB
\newlength\mylenA
\newcommand*\xoverline[2][0.75]{%
    \sbox{\myboxA}{$\m@th#2$}%
    \setbox\myboxB\null
    \ht\myboxB=\ht\myboxA%
    \dp\myboxB=\dp\myboxA%
    \wd\myboxB=#1\wd\myboxA
    \sbox\myboxB{$\m@th\overline{\copy\myboxB}$}
    \setlength\mylenA{\the\wd\myboxA}
    \addtolength\mylenA{-\the\wd\myboxB}%
    \ifdim\wd\myboxB<\wd\myboxA%
       \rlap{\hskip 0.5\mylenA\usebox\myboxB}{\usebox\myboxA}%
    \else
        \hskip -0.5\mylenA\rlap{\usebox\myboxA}{\hskip 0.5\mylenA\usebox\myboxB}%
    \fi}
\makeatother

\def\phi{\varphi}

\def\cC{{\cal C}}

\def\bh{{\mathbf{h}}}

\def\bu{{\mathbf{u}}}
\def\bv{{\mathbf{v}}}

\def\bx{{\mathbf{x}}}
\def\by{{\mathbf{y}}}

\def\b0{{\mathbf{0}}}

\def\bH{{\mathbf{H}}}
\def\bI{{\mathbf{I}}}

\def\bR{{\mathbf{R}}}

\def\cC{\mathcal{C}}

\def\cG{\mathcal{G}}
\def\cH{\mathcal{H}}

\def\cN{\mathcal{N}}

\def \Re[#1]{#1_{\rm R}}
\def \Im[#1]{#1_{\rm I}}
\def \Cpx[#1]{\hat{#1}}
\def \tx {x} 
\def \rx {y} 

\def \rxr {\Re[\rx]}

\def \rxi {\Im[\rx]}

\def \txv {\bx} 
\def \rxv {\by} 

\def \txvr {\Re[\txv]}
\def \rxvr {\Re[\rxv]}
\def \txvi {\Im[\txv]}
\def \rxvi {\Im[\rxv]}
\def \txvc {\Cpx[\txv]}
\def \rxvc {\Cpx[\rxv]}
\def \txvra {\txv_{{\rm R},a}}
\def \txvia {\txv_{{\rm I},a}}

\def \buc {\hat{\bu}}
\def \bvc {\hat{\bv}}
\def \uc {\hat{u}}
\def \vc {\hat{v}}

\def \chm {H} 
\def \chmr {\Re[\chm]}
\def \chmi {\Im[\chm]}
\def \chmc {\Cpx[\chm]}
\def \chvr {\Re[\chv]}
\def \chvi {\Im[\chv]}

\def \snr {\rho}
\def \SNR {{\rm SNR}}

\def \MuI {\bI} 
\def \Ent {\bH} 
\def \Entfuntwo {\cH_2} 

\def \E {\mathbb{E}} 

\def \chv {\bh} 
\def \tn {M} 
\def \rn {N} 
\def \nv {\bv} 

\def \nvr {\Re[\nv]}
\def \nvi {\Im[\nv]}
\def \nvc {\Cpx[\nv]}

\def \Cc {\Cpx[C]}
\def \Cavgc {\Cpx[\Cavg]}
\def \Xic {\Cpx[\Xi]}
\def \cGc {\Cpx[\cG]}

\def \sign {\text{sign}}

\def \ratio {\alpha}
\def \Cavg {\cC} 
\def \cgas {u} 
\def \cgasv {\bu} 

\def \covmtx {R} 
\def \covmtxs {R} 
\def \pdfcov {\mu} 
\def \cgfcov {\Lambda}
\def \tcovmtx {\tilde{\covmtx}}
\def \tcovmtxs {\tilde{\covmtxs}}

\def \Ccrit{\Cavg}

\def \Cbsc {c}

\newcommand{\R}{\mathbb{R}}

\begin{document}

\title{Capacity of Multiple One-Bit Transceivers in a Rayleigh Environment}
\author{\IEEEauthorblockN{Kang Gao, J. Nicholas Laneman, Bertrand Hochwald}\\
\IEEEauthorblockA{Department of Electrical Engineering, University of Notre Dame, Notre Dame, IN, 46556\\
Email: \texttt{\{kgao,jnl,bhochwald\}@nd.edu}}}
\maketitle


\begin{abstract}
We analyze the channel capacity of a system with a large number of one-bit transceivers in a classical Rayleigh environment with perfect channel information at the receiver.  With $\tn$ transmitters and $\rn=\ratio\tn$ receivers, we derive an expression of the capacity per transmitter $\Cavg$, where $\Cavg\leq\min(1,\ratio)$, as a function of $\ratio$ and signal-to-noise ratio (SNR) $\snr$, when $\tn\to\infty$.  We show that our expression is a good approximation for small $\tn$, and provide simple approximations of $\Cavg$ for various ranges of $\alpha$ and $\snr$.  We conclude that at high $\SNR$, $\Cavg$ reaches its upper limit of one only if $\ratio>1.24$.  Expressions for determining when $\Cavg$ ``saturates" as a function of $\ratio$ and $\snr$ are given.
\end{abstract}

\IEEEpeerreviewmaketitle

\section{Introduction}
In an effort to save power and cost in wideband wireless transceiver systems, low-resolution (especially one-bit) analog-to-digital converters (ADCs) \cite{singh2009limits,krone2012capacity,mo2015capacity,mo2014high,mezghani2009analysis,mezghani2007ultra, mezghani2008analysis,mollen2016one,mollen2017uplink,choi2016near,li2016channel,mo2014channel,studer2015quantized} and digital-to-analog converters (DACs) \cite{saxena2016analysis,li2017downlink,jacobsson2017massive} are being considered in transmitter and receiver chains, especially in systems involving many such chains.  The nonlinearity introduced by coarse quantization becomes a limiting factor in the achievable throughput of such a wireless system.  Channel capacity is one measure of this throughput.

There is a rich literature on the subject of capacity with coarse quantization.  The capacity of a system with one-bit ADCs at the receiver is analyzed in \cite{singh2009limits,krone2012capacity,mo2015capacity,mo2014high,mezghani2007ultra, mezghani2008analysis,mezghani2009analysis,mollen2016one,mollen2017uplink} with various assumptions about the channel, the channel information, and communication schemes. Communication techniques including channel estimation and signal detection for a multiple-input multiple-output (MIMO) system with one-bit ADCs at the receiver are studied in \cite{choi2016near,li2016channel,mo2014channel,studer2015quantized}. A communication system with one-bit DACs at the transmitter is studied in \cite{saxena2016analysis,li2017downlink,jacobsson2017massive}.

While many of the efforts consider low-resolution quantization effects at the transmitter or receiver, a few consider low-resolution quantization on both, including \cite{usman2016mmse} and \cite{gao2017power}. A linear minimum-mean-squared-error precoder design is proposed for a downlink massive MIMO scenario to mitigate the quantization distortion in \cite{usman2016mmse} and the performance analysis of a system with a small number of one-bit transceivers is studied in \cite{gao2017power}. We focus on a model where one-bit quantization is considered at both the transmitter and receiver:
\begin{equation}
\rxv=\sign\left(\sqrt{\frac{\snr}{\tn}}\chm\txv+\nv\right), \txv\in\{\pm 1\}^{\tn},
\label{eq:real_channel_model}
\end{equation}
where $\tn$ and $\rn$ are the number of transmitters and receivers, $\txv\in\{\pm1\}^{\tn}$ and $\rxv\in\{\pm1\}^{\rn}$ are the transmitted and received signals, $\chm\in\R^{\rn\times\tn}$ is the channel matrix known to the receiver, $\nv\in\R^{\rn}$ is the additive Gaussian noise with $\nv\sim\cN(0,I)$ and $\nv$ is independent of $\txv$ and $\chm$, $\snr$ is the expected received SNR at each receive antenna. The function $\sign(\cdot)$ provides the sign of the input as its output. The channel is modeled as real-valued since only the in-phase (I) information is used and the quadrature (Q) phase is ignored at the receiver. A Rayleigh channel is assumed, with each element to be independent Gaussian $\cN(0,1)$. This assumption appears to hold for non-line of sight (NLOS) channels in many frequency bands \cite{rappaport2014millimeter}, and also appears in the analysis in \cite{mezghani2007ultra,mezghani2008analysis, mollen2016one,mollen2017uplink,choi2016near,li2016channel,saxena2016analysis}.

Our contribution is a large $\tn$ and $\rn$ analysis, where the ratio $\ratio=\frac{\rn}{\tn}$ is constant, of the capacity for the model \eqref{eq:real_channel_model}.  Analytical expressions are derived that can be used to explain the behavior of the system in various limiting regimes of operation in $\ratio$ and $\SNR$.

\section{Capacity for a large number of transmitters and receivers}
The capacity of the channel in (\ref{eq:real_channel_model}) as a function of $\snr$, $\tn$, and $\rn$ is
\begin{equation}
C(\snr,\tn,\rn) =\frac{1}{\tn} \max\limits_{p_{\rm\tx}(\cdot),\txv\in\{\pm1\}^{\tn}}\MuI(\txv;\rxv,\chm),
\label{eq:capacity_only_rx_know_H_1}
\end{equation}
where we have normalized by $\tn$, and
where $p_{\rm\tx}(\cdot)$ is the input distribution independent of $\chm$, and $\MuI(\cdot;\cdot)$ is the mutual information notation. When $\tn,\rn\to\infty$ with a ratio $\ratio=\frac{\rn}{\tn}$, the capacity $\Cavg$ is defined as
 \begin{equation}
\Cavg(\snr,\ratio) = \lim_{\tn\to\infty}C(\snr,\tn,\ratio\tn).
\label{eq:average_Cap_per_tx}
\end{equation}
We can readily see that $C(\snr,\tn,\rn)\leq\min(1,\ratio)$ because each transmitter can transmit at most one bit of information, and each receiver can decode at most one bit of information.  Therefore, $\Cavg(\snr,\ratio)\leq\min(1,\alpha)$ for all $\ratio$ and $\rho$.

Limiting capacities such as (\ref{eq:average_Cap_per_tx}) are difficult to compute in closed form, but the
``replica method" \cite{mezard1988spin} can be brought to bear on the problem.  Some details of how to apply the method are presented in Section \ref{sec:analysis}.  For now, we present the result:
\begin{align*}
&\Cavg(\snr,\ratio) = \min\Big( \ratio(\Cbsc(\snr)-\Cbsc(A^2q))+\frac{1}{2\ln 2}(E + Eq)\\
&-\frac{1}{\sqrt{2\pi}}\int_{\R}\log_2(\cosh(E+\sqrt{E}z))e^{-z^2/2} dz,1\Big),
\numberthis
\label{eq:Cavg}
\end{align*}
where $\Cbsc(\snr)$ is the capacity of a single transceiver with SNR $\snr$, which is defined as
\begin{equation}
\Cbsc(\snr) = 1-\E_z\left(\Entfuntwo(Q(\sqrt{\snr} z))\right), z\sim\cN(0,1),
\label{eq:single_transceiver_capacity}
\end{equation}
where $\Entfuntwo(p)=-(p\log_2p + (1-p)\log_2(1-p))$ is the binary entropy function, and $q,E,A$ are the solutions of
\begin{equation}
q = \frac{1}{\sqrt{2\pi}}\int_{\R}\tanh(\sqrt{E}z+E)e^{-z^2/2}dz,
\label{eq:q_solution}
\end{equation}
\begin{equation}
E = \frac{\ratio A^2}{\pi\sqrt{2\pi}} \int_{\R}
\frac{\exp\left(-(A^2q+\frac{1}{2})z^2\right)}{Q(A\sqrt{q}z)}dz,
\label{eq:E_solution}
\end{equation}
\begin{equation}
A = \sqrt{\frac{\snr}{1+\snr(1-q)}}.
\label{eq:A_solution}
\end{equation} 

Equation (\ref{eq:Cavg}) gives the capacity for any SNR $\snr$ and $\ratio$, and some limiting situations are readily analyzed, including: (i) high SNR, $\snr\to\infty$; (ii) low SNR, $\snr\to 0$; (iii) many more receivers than transmitters, $\ratio\to\infty$; (iv) many more transmitters than receivers, $\ratio\to 0$.  These are now presented, with only limited proofs.

\subsection{High SNR ($\snr\to \infty$)}
When $\snr\to \infty$, the system is effectively becoming ``noise-free", and we might expect $\Cavg\to 1$, but as we show this does not happen for all $\ratio$.  For $\SNR\to\infty$, 
$A = \sqrt{\frac{1}{1-q}}$, and (\ref{eq:Cavg}) becomes
\begin{align*}
&\Cavg(\ratio,\snr) = \min\Big(\ratio\big(1-\Cbsc(\frac{q}{1-q})\big) + \frac{1}{2\ln 2}(E+Eq)\\
&- \frac{1}{\sqrt{2\pi}}\int_{\R}\log_2(\cosh(E+\sqrt{E}z))e^{-z^2/2} dz,1\Big),
\numberthis
\label{eq:Cavg_H}
\end{align*}
where $\Cbsc(\cdot)$ is defined in (\ref{eq:single_transceiver_capacity}), (\ref{eq:E_solution}) can be simplified as
\begin{equation}
E =  \frac{\ratio}{\pi\sqrt{2\pi(1-q)}} \int_{\R}
\frac{\exp(-(q+\frac{1}{2})z^2)}{Q(\sqrt{q}z)}
dz,
\label{eq:high_SNR_E}
\end{equation}
and $E,q$ are the solution of (\ref{eq:q_solution}) and (\ref{eq:high_SNR_E}).

The expression (\ref{eq:Cavg_H}) is not especially intuitive, but it is not difficult to solve.  We show some numerical examples in Section \ref{sec:numerical}.  It turns out that in this case,
solving (\ref{eq:Cavg_H}) is essentially equivalent to solving for the ``quenched entropy" for Gibbs learning of the Ising perceptron (Section 7.2 in \cite{engel2001statistical}).

\subsection{Low SNR ($\snr\to 0$)}
When $\snr\to 0$,  (\ref{eq:q_solution}), (\ref{eq:E_solution}), and  (\ref{eq:A_solution}) become
\begin{equation}
 q \to \frac{2\ratio\snr}{\pi},\; E \to \frac{2\ratio\snr}{\pi},\; A\to\sqrt{\snr}.
 \label{eq:low_SNR_approx}
\end{equation}

For small $x$, we use a Taylor series expansion to obtain
\begin{align*}
Q(x)&\ln(Q(x)) = -\frac{\ln2}{2}-\frac{(1-\ln2)x}{\sqrt{2\pi}} + \frac{x^2}{2\pi}\\
 &+ \frac{(1-\ln 2)\pi+2\sqrt{2\pi}}{6\pi\sqrt{2\pi}}x^3+\frac{\pi-1}{6\pi^2}x^4+o(x^4),\\
\ln&(\cosh(x)) = \frac{x^2}{2}-\frac{x^4}{12}+o(x^4).
\end{align*}
Then
\begin{equation}
\Cavg(\snr,\ratio)\approx \frac{\ratio\snr}{\pi\ln2} -\frac{\ratio^2+(\pi-1)\ratio}{\pi^2\ln2}\snr^2,
\label{eq:low_snr_slope}
\end{equation}
to second order in $\snr$.  It turns out that this result matches the expression in \cite{mezghani2007ultra}, with a difference in factor of $\frac{1}{2\ln2}$ that comes from the fact that \cite{mezghani2007ultra} considers two bits per transmission, and `nats' instead of `bits'.

\subsection{$\rn>>\tn$ ($\ratio\to\infty$)}
When $\ratio\to\infty$, (\ref{eq:q_solution})-(\ref{eq:A_solution}) becomes
\begin{equation*}
q \to 1, A\to\sqrt{\snr}, 
\end{equation*}
\begin{equation}
E\to \frac{\ratio\snr}{\pi\sqrt{2\pi}}\int_{\R}\frac{\exp\left(-(\snr+\frac{1}{2}) z^2\right)}{Q(\sqrt{\snr}z)}dz.
\label{eq:large_alpha_saddle_point}
\end{equation}
and
\begin{align*}
&\Cavg(\snr,\ratio) \approx \min\Big(1, \frac{E}{\ln 2} - \\
&\int_{\R} \frac{1}{\sqrt{2\pi}}e^{-z^2/2}\log_2(\cosh(E+\sqrt{E}z))dz\Big),
\numberthis
\label{eq:large_alpha_slope_result}
\end{align*}
More will be said about this in Section \ref{sec:numerical}.

\subsection{$\rn<<\tn$ ($\ratio\to 0$)}
When $\ratio\to 0$, the first order approximations of $q$ and $E$ based on (\ref{eq:q_solution})-(\ref{eq:A_solution}) become
\begin{equation}
q = \frac{2\snr\ratio}{(1+\snr)\pi}+o(\ratio),
E = \frac{2\snr\ratio}{(1+\snr)\pi}+o(\ratio).
\end{equation}
Therefore,
\begin{equation}
\Cavg(\snr,\ratio)\approx\Cbsc(\snr)\ratio - \frac{\snr^2}{\pi^2(1+\snr)^2\ln 2}\ratio^2,
\label{eq:more_TX_slope}
\end{equation}
where $\Cbsc(\snr)$ is the capacity of a single transceiver, defined in (\ref{eq:single_transceiver_capacity}).

\section{numerical evaluation of capacity}
\label{sec:numerical}

\begin{figure}
\includegraphics[width=3.4in]{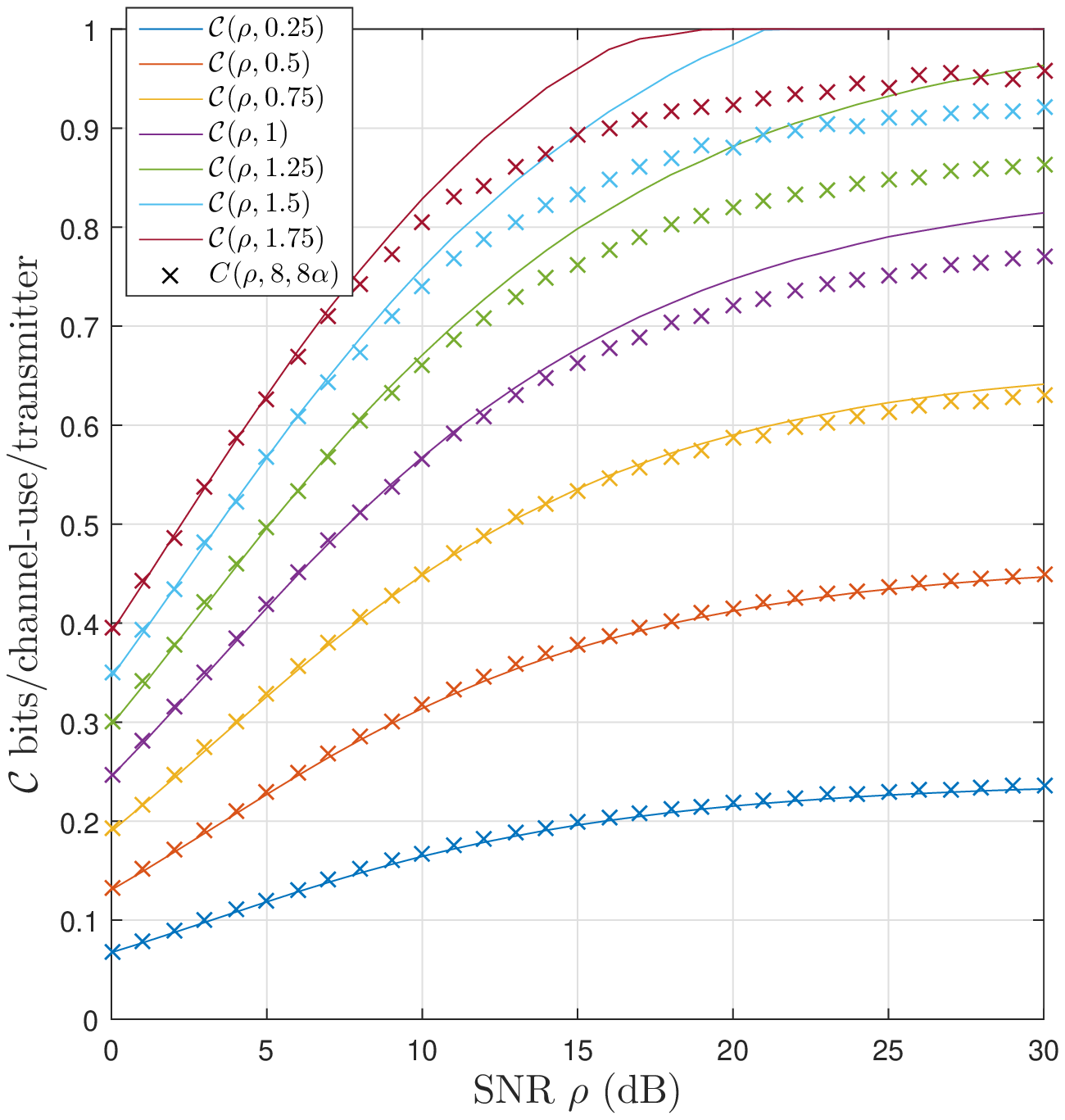}
\centering
    \caption{Comparison between $C(\snr,\tn,\ratio\tn)$ (\ref{eq:capacity_only_rx_know_H_1})  and $\Cavg(\snr,\ratio)$ (\ref{eq:Cavg}) with $\tn=8$ for $\ratio\in\{0.25,\cdots,1.75\}$ with SNR varying from 0 dB to 30 dB. The accuracy of the approximation in a wide range of SNR shows that $\tn=8$ is big enough to get a valid approximation with $\Cavg(\snr,\ratio)\leq 0.7$.}
    \label{fig:C_over_M_and_RS_comparison}
\end{figure}

We first compare $C(\snr,\tn,\ratio\tn)$ in (\ref{eq:capacity_only_rx_know_H_1}) with $\cC(\snr,\ratio)$ in (\ref{eq:Cavg}) for $\tn=8$ and $\ratio\in\{0.25,\cdots,1.75\}$, to show how the large-$M$ and $N$ limit (\ref{eq:Cavg}) can be used to approximate the exact capacity.  Figure \ref{fig:C_over_M_and_RS_comparison} displays that the approximation is accurate for small $\tn$ for a wide range of SNR (from 0 dB to 30 dB) when $\Cavg(\snr,\ratio)\leq 0.7$. When $\Cavg(\snr,\ratio)$ is larger than 0.7, $\tn=8$ is not big enough and a larger $\tn$ is required to get a valid approximation. We can see that $\Cavg(\snr,\ratio)$ can saturate at 1 when $\ratio\geq1.5$ with SNR smaller than 30 dB, and an SNR higher than 30 dB is required to achieve the maximum for $\ratio=1.25$, but $\Cavg(\snr,\ratio)$ cannot achieve the maximum when $\ratio\leq 1$. We will show later that $\ratio>1.24$ is required to achieve the maximum.  

\begin{figure}
\includegraphics[width=3.4in]{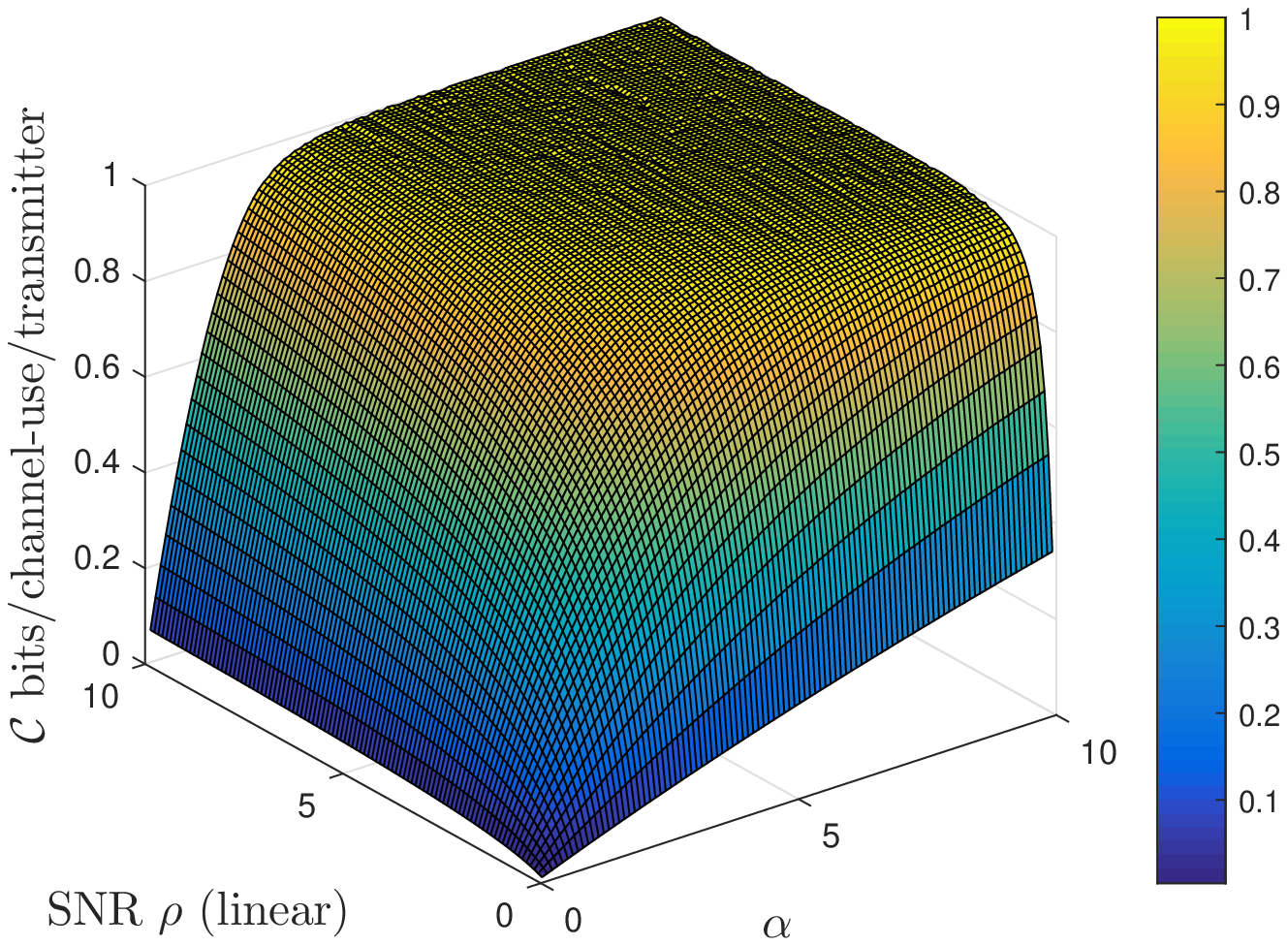}
\centering
    \caption{$\Cavg(\snr,\ratio)$ (\ref{eq:Cavg}) versus $\snr$ and $\ratio$ with $0.1\leq\snr\leq 10$ and $0.1\leq\ratio\leq 10$. $\Cavg(\snr,\ratio)$ increases linearly with $\snr$ and $\ratio$ when $\snr$ and $\ratio$ are small but $\Cavg(\snr,\ratio)$ ``saturates" quickly.} 
    \label{fig:comparision_table_C_SNR_alpha}
\end{figure}

 Figure \ref{fig:comparision_table_C_SNR_alpha} displays $\Cavg(\snr,\ratio)$ in (\ref{eq:Cavg}) for $\ratio$ and $\snr$ varying from 0.1 to 10 with step 0.1. We can see that $\Cavg(\snr,\ratio)$ increases linearly with $\snr$ and $\ratio$ when $\snr$ and $\ratio$ are small, and the rate of increase slows down dramatically as $\snr$ and $\ratio$ grow and $\Cavg$ nears saturation at $\Cavg\approx 1$. When $\ratio$ is small but $\snr$ is large, $\Cavg(\snr,\ratio)$ saturates at $\Cavg(\snr,\ratio)\approx\ratio$ (its upper bound). When $\snr$ is small but $\ratio$ is large, we can see that $\Cavg(\snr,\ratio)$ increases with $\ratio$ and reaches its maximum value 1. We show that for any $\snr$, $\Cavg(\snr,\ratio)\to 1$ when $\ratio\to\infty$.
 
\begin{figure}   
\includegraphics[width=3.4in]{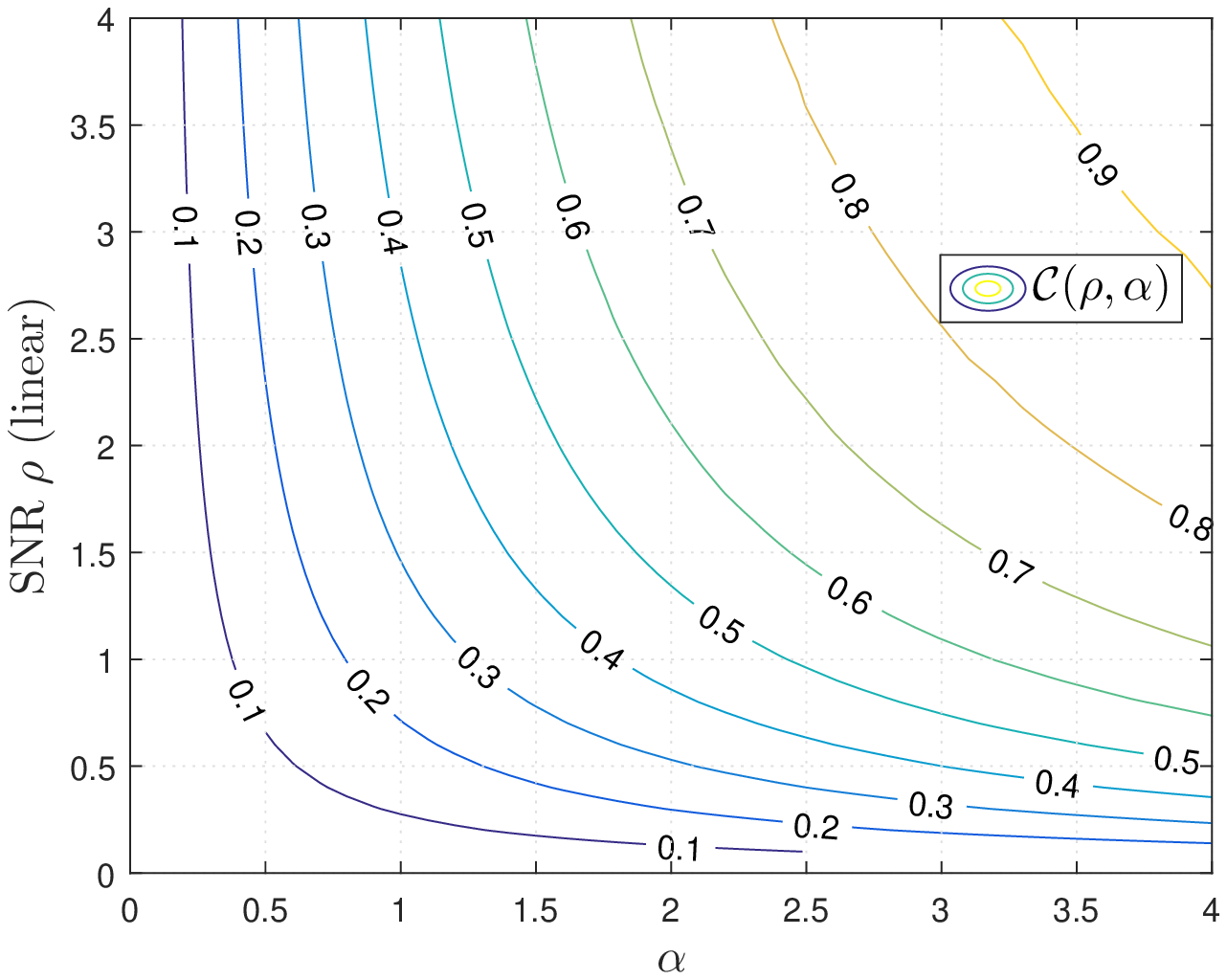}
\centering
    \caption{Contour plot of $\Cavg(\snr,\ratio)$ (\ref{eq:Cavg}) with $\ratio\leq4$ and $\snr\leq4$. The gap between the adjacent contours becomes large as $\Cavg(\snr,\ratio)$ increases, which shows the increasing demands on $\snr$ and $\ratio$ to increase $\Cavg(\snr,\ratio)$.}
    \label{fig:contour_C_vs_snr_alpha}
\end{figure}

 Contours of constant $\Cavg(\snr,\ratio)$ for $\snr \leq 4$ and $\ratio \leq 4$ are shown in Figure \ref{fig:contour_C_vs_snr_alpha}.  We can observe that there is generally a sharp tradeoff between $\snr$ and $\ratio$, and that operating near the knee in the curve is generally desirable for a given capacity since both $\ratio$ and $\snr$ are small.
 
 Furthermore, the contours are dense when $\Cavg(\snr,\ratio)\leq 0.8$ and start becoming sparse when $\Cavg(\snr,\ratio)\geq 0.8$, thus showing that $\Cavg(\snr,\ratio)$ has started to ``saturate" at 0.8 and improves only slowly with further increases in either $\ratio$ or $\snr$.  
 
 The contours allow us to explore optimal operating points.  For example, given a cost function where $\ratio+\snr=d$ for some constant $d$, we find an approximately optimal point to achieve $\Cavg(\snr,\ratio)=0.8$ is $\ratio=3.4$ and $\snr\approx 2.07$.  Attempts to make $\ratio$ smaller will require significant increase in $\snr$, and attempts to make $\snr$ smaller will require significant increase in $\ratio$.

\begin{figure}
\includegraphics[width=3.2in]{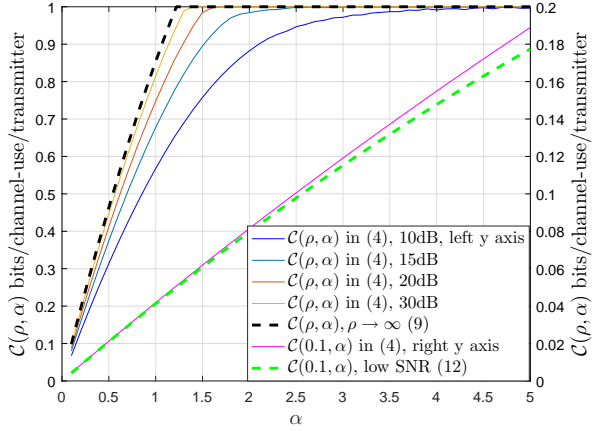}
\centering
    \caption{Comparison between $\Cavg(\snr,\ratio)$ (\ref{eq:Cavg}) at high SNR (10 dB to 30 dB), low SNR (-10 dB, $\snr = 0.1$), and their corresponding approximations in (\ref{eq:Cavg_H}) and (\ref{eq:low_snr_slope}). The curve at $\snr=10$ (10 dB) is already close to the noise-free case, which increases nearly linearly with $\ratio$ before saturation at $\ratio\approx1.24$. The low SNR approximation (\ref{eq:low_snr_slope}) is similarly accurate when $\snr = 0.1$ for $\ratio\leq 4$. In general, we need $\ratio\leq 0.4/\snr$ for an accurate low SNR approximation when $\snr\leq 0.1$.}
    \label{fig:large_small_SNR_comparison}
\end{figure}

Figure \ref{fig:large_small_SNR_comparison} shows the accuracy of the approximations of $\Cavg(\snr,\ratio)$ at high and low SNR.  Plotted are examples when SNR is large (10 dB to 30 dB) and SNR is low (-10 dB, $\snr=0.1$) of the actual capacity (\ref{eq:Cavg}) and the corresponding approximations (\ref{eq:Cavg_H}) and (\ref{eq:low_snr_slope}).  Of particular interest is the approximately linear growth in (\ref{eq:Cavg_H}) with $\ratio$ until it reaches the $\Cavg=1$ saturation point when $\ratio\approx 1.24$.  This hard limit value of 1.24 receive antennas for every transmit antenna is perhaps surprising.

The curves for low SNR show that $\Cavg(\snr,\ratio)$ at $\snr=0.1$ is close to the low SNR approximation in (\ref{eq:low_snr_slope}) with a simple second order expression when $\ratio\leq 4$. In general, when $\snr\leq 0.1$, we need $\ratio\leq 0.4/\snr$ for an accurate low SNR approximation according to (\ref{eq:low_SNR_approx}).

\begin{figure}
\includegraphics[width=3.2in]{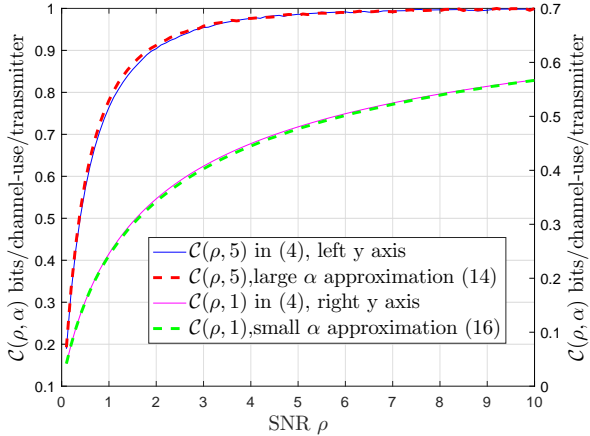}
\centering
    \caption{The comparison between the capacity $\Cavg(\snr,\ratio)$ (\ref{eq:Cavg}) with large $\ratio$ ($\ratio=5$), small $\ratio$ ($\ratio=1$) and their corresponding approximations in (\ref{eq:large_alpha_slope_result}) and (\ref{eq:more_TX_slope}). The good approximations over a wide range of SNR shows that $\ratio=5$ is big enough and $\ratio=1$ is small enough to use (\ref{eq:large_alpha_slope_result}) and (\ref{eq:more_TX_slope}) for accurate approximations.}
    \label{fig:large_small_alpha_comparison}
\end{figure}

Figure \ref{fig:large_small_alpha_comparison} presents a comparison of (\ref{eq:Cavg}) with the large $\ratio$, small $\ratio$ approximations in (\ref{eq:large_alpha_slope_result}) and (\ref{eq:more_TX_slope}). We obtain excellent agreement for even the modest values $\ratio=5$ and $\ratio=1$ over a wide range of SNR. Moreover, according to (\ref{eq:large_alpha_saddle_point}), when $\ratio\to\infty$, we have $E\to\infty$ for any $\snr>0$, and thus $\Cavg(\snr,\ratio)\to 1$. This differs from the high SNR case, where $\Cavg(\snr,\ratio)<1$ when $\ratio<1.24$ even as $\snr\to\infty$.

\subsection{Tradeoff between $\ratio$ and $\snr$ for fixed $\Cavg$}

We are interested in characterizing the contours in Figure \ref{fig:contour_C_vs_snr_alpha} analytically, and we use the large $\ratio$ approximation for $\Cavg$ in (\ref{eq:large_alpha_slope_result}).  Since $\Cavg$ in (\ref{eq:large_alpha_slope_result}) is just a function of $E$, to achieve some target capacity $\Ccrit$, we solve for $E$ numerically, and denote the result as $E_{\Ccrit}$.  With $E=E_{\Ccrit}$, (\ref{eq:large_alpha_saddle_point}) then provides the relationship between $\ratio$ and $\snr$. To simplify the relationship, (\ref{eq:large_alpha_saddle_point}) can be further approximated as
\begin{equation}
E_{\Ccrit}\approx\frac{\ratio}{\pi}(-0.3\snr^2+1.8\snr)
\label{eq:E_approx_high_SNR}
\end{equation}
with good accuracy when $\snr\leq 1.5$. The relationship between $\ratio$ and $\snr$ is then
\begin{equation}
\snr \approx 3-\sqrt{9 - \frac{10E_{\Ccrit}\pi}{3\ratio}}.
\label{eq:SNR_requirement_large_ratio}
\end{equation}

\begin{figure}
\includegraphics[width=3.4in]{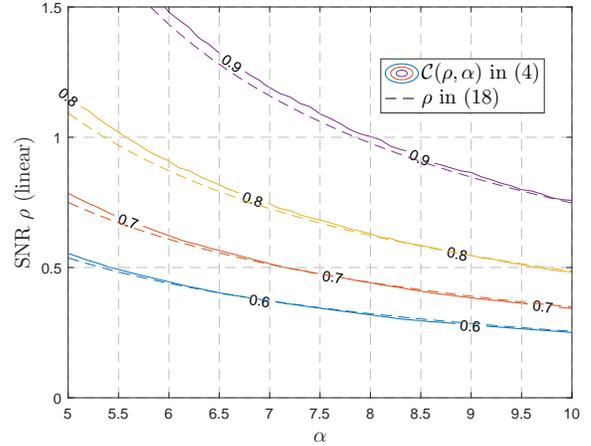}
\centering
    \caption{Comparison between the actual SNR $\snr$ and the approximated SNR (\ref{eq:SNR_requirement_large_ratio}) for the contours $\Cavg=0.6,\ldots,0.9$. The solid lines are contour plots of (\ref{eq:Cavg}), and the dashed lines present the approximation of $\snr$ (\ref{eq:SNR_requirement_large_ratio}) as a function of $\ratio$.} 
    \label{fig:large_alpha_approx_SNR_vs_true_SNR}
\end{figure}

To verify the approximation in (\ref{eq:SNR_requirement_large_ratio}), we compare the actual SNR $\snr$ with the approximated $\snr$ (\ref{eq:SNR_requirement_large_ratio}) in Figure \ref{fig:large_alpha_approx_SNR_vs_true_SNR}.  Shown are contours for $\Ccrit=0.6,\ldots,0.9$ and $\ratio\in[5,10]$. The solid lines are the contour plot of $\Cavg(\snr,\ratio)$ (\ref{eq:Cavg}), and the dashed lines are (\ref{eq:SNR_requirement_large_ratio}).  We see good agreement over a wide range of $\ratio$.  

Clearly, there are many other comparisons and tradeoffs we can analyze using (\ref{eq:Cavg}) and its approximations.  We briefly describe how the replica method was applied to obtain (\ref{eq:Cavg}).

\section{Replica Analysis}
\label{sec:analysis}

The replica method, a tool used in statistical mechanics and applied to the theory of spin glass \cite{mezard1988spin}, has been applied in many communication system contexts \cite{tanaka2001analysis,tanaka2002statistical,tanaka2004statistical,guo2005randomly,guo2003multiuser}, neural networks \cite{gardner1988space,engel2001statistical}, error-correcting codes \cite{montanari2000statistical}, and image restoration \cite{nishimori1999statistical}.  A mathematically rigorous justification of the replica method is elusive, but the success of the method maintains its popularity. We apply the method to solve for a closed-form answer to (\ref{eq:average_Cap_per_tx}).  We omit many details, and present only the primary steps.

Because the channel is unknown to the transmitter, according to \cite{gao2017power}, the optimal input distribution is $p_{\tx}(\txv)=\frac{1}{2^{\tn}}$, and then (\ref{eq:capacity_only_rx_know_H_1}) becomes
\begin{align*}
C(\snr,\tn,\rn)&=\frac{1}{\tn} \left(\Ent(\rxv|\chm) - \Ent(\rxv|\txv,\chm)\right), p_{\tx}(\txv)=\frac{1}{2^{\tn}},
\end{align*}
where $\Ent(\cdot)$ is the standard definition of entropy.

Since the elements in $\chm$ are i.i.d. $\cN(0,1)$, and $\txv\in\{\pm1\}^{\tn}$, we have $\sqrt{\frac{\snr}{\tn}}\chm\txv\sim\cN(0,\snr I)$, and
\begin{equation}
\Ent(\rxv|\txv,\chm) = \rn (1-\Cbsc(\snr)),
\label{eq:entropy_condition_x_H}
\end{equation}
where $c(\snr)$ is defined in (\ref{eq:single_transceiver_capacity}).

Then, (\ref{eq:average_Cap_per_tx}) becomes
\begin{equation}
\Cavg(\snr,\ratio) = \lim_{\tn\to\infty}\frac{\Ent(\rxv|\chm)}{\tn} - \ratio (1-\Cbsc(\snr)).
\label{eq:Cavg1}
\end{equation}
The replica method is used to compute the limit, and the processes are similar to that used in \cite{tanaka2001analysis,tanaka2002statistical,tanaka2004statistical,guo2005randomly,guo2003multiuser}. 

We start with the identity 
\begin{align*}
\Ent(\rxv|\chm) = -\frac{1}{\ln 2}\lim_{n \to 0}\frac{\partial}{\partial n} \ln \E_{\chm,\rxv} (p(\rxv|\chm))^n,
\end{align*}
which holds for $n\in \R$. The idea of the replica method is to compute $\E_{\chm,\rxv} (p(\rxv|\chm))^n$ as a function of $n$ by treating $n$ as a positive integer, and then assume the result to be valid for $n \in \R$.

We assume the limit of $\tn$ and $n$ can commute, then
\begin{equation}
    \underset{\tn\to\infty}{\lim}\frac{\Ent(\rxv|\chm)}{\tn} = -\frac{1}{\ln 2}\lim_{n\to 0}\frac{\partial}{\partial n}\lim_{\tn \to\infty} \frac{\ln \Xi_n}{\tn},
    \label{eq:entropy_rxv_avg}
\end{equation}
where $\Xi_n=\E_{\chm,\rxv} (p(\rxv|\chm))^n$.

Now, we regard $n$ as a positive integer and we have
\begin{align*}
    \Xi_n &= \E_{\chm,\rxv} (p(\rxv|\chm))^n = \E_{\chm}\sum_{\rxv}\left(\prod_{a=0}^{n}\E_{\txv_a} p(\rxv|\chm,\txv_a) \right),
\end{align*}
where $\txv_a$ is the $a$th replica of $\txv$ ($0\leq a \leq n$), and $\txv_a$ are i.i.d. uniform distributed in $\{\pm1\}^{\tn}$.

Based on (\ref{eq:real_channel_model}), we further have
\begin{align*}
    \Xi_n &= \E_{\txv_0,\cdots,\txv_n}\sum_{\rxv}\E_{\chm}\left(
    \prod_{a=0}^{n} \prod_{k=1}^{{\rn}}Q\left(-\rx_k\sqrt{\frac{\snr}{\tn}}\chv_k^T\txv_a\right)
    \right)\\
    &= \E_{\txv_0,\cdots,\txv_n} [e^{\rn \cG}],
    \numberthis
    \label{eq:Xi_exp}
\end{align*}
where $\rx_k$ is the $k$th element of $\rxv$, $Q(\cdot)$ is the well-known Q-function, and $\chv_k^T$ is the $k$th row of $\chm$,
\begin{equation}
e^{\cG} = \sum_{\rx}\E_{\chv}
    \prod_{a=0}^{n}Q\left(-\rx\sqrt{\frac{\snr}{\tn}}\chv^T\txv_a\right)
\end{equation}
with $\rx\in\{\pm1\}, \chv\sim\cN(0,I)$.

Let $\cgas_a = \frac{1}{\tn}\chv^T\txv_a$ and $\cgasv=[\cgas_0,\cdots,\cgas_n]^T$. Then $\cgasv\sim\cN(0,\covmtx)$, where $\covmtx$ is the covariance matrix with elements $\covmtxs_{ab}=\E(\cgas_a\cgas_b)=\frac{\txv_a^T\txv_b}{\tn}\in[-1,1]$ for $0\leq a < b \leq n$ and $\covmtxs_{aa}=1$.  Then $\cG$ only depends on $\covmtx$:
\begin{equation}
\cG(\covmtx) = \ln\left(\sum_{\rx}\E_{\cgasv}
    \prod_{a=0}^{n}Q\left(-\rx\sqrt{\snr}\cgas_a\right)\right),
    \label{eq:cG_value}
\end{equation}
and (\ref{eq:Xi_exp}) becomes
\begin{align*}
    \Xi_n  = \int_{\R}\cdots\int_{\R} \prod_{a<b}d\covmtxs_{ab}\pdfcov_{\tn}(\covmtx)e^{\rn \cG(\covmtx)},
\end{align*}
where
\begin{equation}
\pdfcov_{\tn}(\covmtx)=\E_{\txv_0,\cdots,\txv_n}\left(\prod_{a<b }\delta(\frac{\txv_a^T\txv_b}{\tn}-\covmtxs_{ab})\right).
\label{eq:pdf_R}
\end{equation}

We can consider $\pdfcov_{\tn}(\covmtx)$ as the distribution of a random symmetric matrix $\bR$, and we have
\begin{align*}
    \Xi_n  = \E_{\bR}\Big[e^{\rn \cG(\bR)}\Big],\bR\sim\pdfcov_{\tn}(\covmtx)
\end{align*}

Similarly to \cite{tanaka2004statistical}, we apply Varadhan's theorem and Gartner-Ellis theorem\cite{dembo38large} and obtain
\begin{equation}
\lim_{\tn \to\infty} \frac{\ln \Xi_n}{\tn} = \sup_{\covmtx}\inf_{\tcovmtx}\underbrace{\left[
\ratio \cG(\covmtx) - \sum_{a<b}\tcovmtxs_{ab}\covmtxs_{ab} + \cgfcov(\tcovmtxs)\right]}_{f(\covmtx,\tcovmtx)},
\label{eq:Replica_Saddle}
\end{equation}
\topskip=0.2in
where $\tcovmtx$ is an $(n+1)\times(n+1)$ matrix with $\tcovmtxs_{ab}$ as its elements, and $\cgfcov(\tcovmtxs)$ is defined as
\begin{align}
\cgfcov(\tcovmtxs) = \lim_{\tn\to \infty}\frac{1}{\tn}\ln\E_{\bR}\exp\left(\sum_{a<b}\tn\tcovmtxs_{ab}\bR_{ab}\right).
\end{align}
Based on the distribution of $\bR$ in (\ref{eq:pdf_R}), we further have
\begin{align*}
  &\cgfcov(\tcovmtxs)  = 
 \ln \left[\E_{\tx_0,\cdots,\tx_n} \exp\left( \sum_{a<b} \tcovmtxs_{ab}\tx_a\tx_b
 \right) \right],
 \numberthis
 \label{eq:cgf_value}
\end{align*}
where $\tx_a$ are independent uniform distributed in $\{\pm1\}$.

$\covmtx$ and $\tcovmtx$ that achieve the optimal value described in (\ref{eq:Replica_Saddle}) are called the saddle point\cite{engel2001statistical}. The saddle point either stays on the boundary ($\covmtxs_{ab}=1$ or $\covmtxs_{ab}=-1$) or satisfies $\frac{\partial f}{\partial \tcovmtxs_{ab}} = \frac{\partial f}{\partial \covmtxs_{ab}} = 0$, i.e.
\begin{equation}
\covmtxs_{ab}=\frac{\partial \cgfcov(\tcovmtxs)}{\partial \tcovmtxs_{ab}}, \tcovmtxs_{ab}=\ratio \frac{\partial \cG(\covmtx)}{\partial \covmtxs_{ab}}, (a<b)
\label{eq:saddle_point}
\end{equation}
with $\cG(\covmtx)$ and $\cgfcov(\tcovmtxs)$ expressed in (\ref{eq:cG_value}) and (\ref{eq:cgf_value}).

Here, we further assume that permutations among the $(n+1)$ replicas with index $a=0,1,2,\cdots,n$ will not change the saddle point. This assumption is called the ``replica symmetry" (RS) assumption.
At the saddle point, we let
\begin{equation}
\covmtxs_{ab}=q, \tcovmtxs_{ab}=E, (0\leq a < b \leq n),
\label{eq:RS_assumption}
\end{equation}
which are called RS saddle points.

Based on (\ref{eq:Cavg1}), (\ref{eq:entropy_rxv_avg}), (\ref{eq:Replica_Saddle}), and (\ref{eq:RS_assumption}), we can get (\ref{eq:Cavg}), where $\Cavg(\snr,\ratio)=1$ is the solution when the saddle point is on the boundary ($q=1$).  The remaining expressions in (\ref{eq:Cavg}) are obtained when the saddle point satisfies (\ref{eq:saddle_point}), and the corresponding RS saddle point is the solution of (\ref{eq:q_solution})-(\ref{eq:A_solution}).

\subsection{Extension to complex signals}
The real-valued model \eqref{eq:real_channel_model} is now extended to both
I and Q phase at the transmitter and receiver, and hence
\begin{equation}
\rxvc =\sign\left(\sqrt{\frac{\snr}{2\tn}}
\chmc\txvc + \nvc\right), \txvc\in\{\pm 1\}^{2\tn}
\label{eq:complex_channel_model}
\end{equation}
where $\txvc, \rxvc, \chmc, \nvc$ are defined as
\begin{equation*}
\txvc = \begin{bmatrix}\txvr \\ \txvi \end{bmatrix},
\rxvc = \begin{bmatrix}\rxvr \\ \rxvi \end{bmatrix}, 
\chmc = \begin{bmatrix}\chmr & -\chmi\\ \chmi & \chmr \end{bmatrix},
\nvc = \begin{bmatrix}\nvr \\ \nvi \end{bmatrix}.
\end{equation*}
where $\chmr,\chmi\in\R^{\rn\times\tn}$ are the real and imaginary parts of the channel, $\txvr,\txvi\in\{\pm 1\}^{\tn}$ and $\rxvr,\rxvi\in\{\pm 1\}^{\rn}$ are the real and imaginary parts of the transmitted and received signal, and $\nvr$ and $\nvi$ are additive noise. The elements in $\chmr$, $\chmi$, $\nvr$, and $\nvi$ are independent Gaussian $\cN(0,1)$, and $\snr$ is the expected received SNR at each receive antenna.

Since the channel is known only to the receiver, the uniform input is optimal and the channel capacity is
\begin{align*}
\Cc(\snr,\tn,\rn)&=\frac{1}{\tn} \left(\Ent(\rxvc|\chmc) - \Ent(\rxvc|\txvc,\chmc)\right), p_{\hat{\rm\tx}}(\txvc)=\frac{1}{2^{2\tn}}.
\end{align*}
\topskip=0.2in
When $\tn,\rn\to\infty$ with a ratio $\ratio=\frac{\rn}{\tn}$, the capacity $\Cavgc$ is defined as
 \begin{equation}
\Cavgc(\snr,\ratio) = \lim_{\tn\to\infty}\Cc(\snr,\tn,\ratio\tn).
\label{eq:average_Cap_per_tx_complex}
\end{equation}

Similarly to the analysis for real signal, we have
\begin{equation}
\Ent(\rxvc|\txvc,\chmc) = 2\rn (1-\Cbsc(\snr)),
\label{eq:entropy_condition_x_H_c}
\end{equation}
\begin{equation}
\Cavgc(\snr,\ratio) = \lim_{\tn\to\infty}\frac{\Ent(\rxvc|\chmc)}{\tn} - 2\ratio (1-\Cbsc(\snr)).
\label{eq:Cavg1_c}
\end{equation}
Using the replica method with the RS assumption, we obtain
\begin{equation}
    \underset{\tn\to\infty}{\lim}\frac{\Ent(\rxvc|\chmc)}{\tn} = 2\underset{\tn\to\infty}{\lim}\frac{\Ent(\rxv|\chm)}{\tn},
\end{equation}
and therefore
 \begin{equation}
\Cavgc(\snr,\ratio) = 2\Cavg(\snr,\ratio).
\label{eq:average_Cap_per_tx_compare}
\end{equation}
Consequently, the I-Q model capacity is twice the I-only capacity.

\section{Conclusion}
We have presented the capacity per transmitter in the limit where the number of single-bit transmitters $\tn$ and receivers $\rn$ is large, and where $\ratio=\rn/\tn$ was fixed.  A flat Rayleigh fading channel was considered, and we assumed the channel was only known by the receiver.  We were able to derive a variety of approximations using the analytical results, and showed that the large-system formulas are useful even for a small numbers of transmitters and receivers.  We examined how $\Cavg$ saturated with either large $\ratio$ or $\rho$, and gave formulas for exploring the contours of fixed $\Cavg$ as a function of $\ratio$ and $\snr$.  Further work in expanding these results to different channel models would be of great interest.

\section*{Acknowledgment}

The authors are grateful for the support of the National Science Foundation, grants ECCS-1731056, ECCS-1509188, and CCF-1403458.

\bibliographystyle{IEEEtran}
\bibliography{IEEEabrv,WCNC_ref}

\end{document}